

\font\llbf=cmbx10 scaled\magstep2
\font\lbf=cmbx10 scaled\magstep1

\def\index{\hbox{\rm index}\hskip 3pt}
\def\supp{\hbox{\rm supp}\hskip 3pt}
\def\const{\hbox{\rm const}\hskip 3pt}
\def\Area{\hbox{\rm Area}\hskip 3pt}

\def\thorn{ { \rceil \kern -6pt \supset }}
\def\edth{ \hskip 3pt {}^{\prime } \kern -6pt \partial }

\def\D{\hskip 3pt / \kern -7pt D}
\def\uA{\underline A \,}
\def\uB{\underline B \,}
\def\uC{\underline C \,}
\def\uD{\underline D \,}

\hsize=15truecm
\vsize=22truecm
\parindent=1truecm
\baselineskip 14pt plus 2pt

\centerline{\llbf Two dimensional Sen connections and}\par
\centerline{\llbf quasi-local energy-momentum}\par
\vskip 0.5truecm
\centerline{\bf L. B. Szabados}\par
\centerline{Research Institute for Particle and Nuclear Physics}\par
\centerline{H--1525 Budapest 114, P.O.Box 49, Hungary}\par
\centerline{E-mail: lbszab@rmki.kfki.hu}\par
\vskip 1truecm

\centerline{\bf Abstract}\par
\noindent The recently constructed two dimensional Sen connection is
applied in the problem of quasi-local energy-momentum in general
relativity. First it is shown that, because of one of the two 2
dimensional Sen--Witten identities, Penrose's quasi-local charge integral
can be expressed as a Nester--Witten integral. Then, to find the
appropriate spinor propagation laws to the Nester--Witten integral, all
the possible first order linear differential operators that can be
constructed only from the irreducible chiral parts of the Sen operator
alone are determined and examined. It is only the holomorphy or
anti-holomorphy operator that can define acceptable propagation laws.
The 2 dimensional Sen connection thus naturally defines a quasi-local
energy-momentum, which is precisely that of Dougan and Mason. Then
provided the dominant energy condition holds and the 2-sphere is convex
we show that the following statements are equivalent: i. the quasi-local
mass (energy-momentum) associated with a 2-sphere $\$ $ is zero; ii. the
Cauchy development $D(\Sigma)$ is a {\it pp}-wave geometry with pure
radiation ($D(\Sigma)$ is flat), where $\Sigma$ is a spacelike
hypersurface with $\partial\Sigma=\$ $; iii. there exist a Sen--constant
spinor field (two spinor fields) on $\$ $. Thus the {\it pp}-wave Cauchy
developments can be characterized by the geometry of a {\it two} rather
than a {\it three} dimensional submanifold.\par

\vskip 1.5truecm

{\lbf Introduction}\par \vskip 0.5truecm

\noindent This paper is the second part of a four part series on the
theory and applications of the two dimensional Sen connection in
general relativity. In the first of this series [1] a covariant spinor
formalism was developed which is the two dimensional version of the usual
(three dimensional) Sen connection. As the first application of this
formalism quasi-local energy-momentum expressions based on the spinorial
Nester--Witten 2-form will be examined [2-6]. The different constructions
correspond to different additional spinor propagation laws within the
2-surface, and the solutions of these spinor equations are interpreted as
the spinor constituents of the `quasi-translations' of the 2-surface. The
question is therefore how to define the `quasi-translations' of the
2-surface $\$ $. If $\$ $ is in the Minkowski spacetime then the
`quasi-translations' can be expected to coincide with the familiar
translational Killing vectors at the points of $\$ $. To ensure this
coincidence the spinor propagation laws must contain some extrinsic
geometrical properties of $\$ $. The usual formalism in the spinorial
approaches of the quasi-local energy-momentum is the GHP formalism [7-10].
In the GHP formalism the two edth operators, $\edth$ and $\edth^\prime$,
are the covariant directional derivations with respect to the
induced {\it intrinsic} Levi-Civit\`a connection; and the extrinsic
curvatures of $\$ $ are encoded into the spin coeffitients $\rho$, $\sigma$
and $\rho^\prime$, $\sigma^\prime$. The thorn operators, $\thorn$ and
$\thorn^\prime$, and the remaining spin coeffitients $\kappa$, $\tau$
and $\kappa^\prime$, $\tau^\prime$ all depend not only on the geometry of
$\$ $ but the way how the normals to $\$ $ are extended off $\$ $. The
quasi-local energy-momentum, however, is expected to depend only on the
(intrinsic and extrinsic) geometry of $\$ $. Thus in constructing the
propagation laws we can use only the operators $\edth$ and $\edth^\prime$,
the spin coeffitients $\sigma$, $\rho$ and $\sigma^\prime$, $\rho^\prime$
and possibly those components of the curvature that are determined by
$\$ $. In spite of these restriction there remain too much freedom to
construct the propagation laws. The two dimensional Sen connection, on the
other hand, contains all the information on the extrinsic geometry of $\$ $
since the Sen operator is the sum of the intrinsic covariant derivation
and the (boost-gauge invariant combination of the) extrinsic curvatures [1].
Thus the 2 dimensional Sen operator alone might be, and as we will show,
is enough to construct the spinor propagation laws. \par
     In the first section of the present paper the 2-surface integral of
the spinorial Nester--Witten 2-form will be examined. As a consequence of
the 2 dimensional Sen--Witten identities Penrose's construction [2] can
also be considered as a Nester--Witten integral. The possible propagation
laws will be considered and discussed in section 2. All the possible first
order differential operators, acting on the covariant spinor fields, that
can be constructed out of the the chiral irreducible parts of the 2
dimensional Sen operator will be determined. In particular, the properties
of the holomorphy/anti-holomorphy operators will be examined in detail.
Here we use the GHP form of the chiral irreducible parts of the Sen
operator. To clarify the kernel space of the possible first order operators
we need to know the structure of the kernel spaces of the edth operators.
This clarification is made in the Appendix for 2-surfaces homeomorphic to
$S^2$, using only Liouville's theorem and Baston's formula [10] for the
analitic index of $\edth$ and $\edth^\prime$. We will show that the Sen
operator naturally defines a quasi-local energy-momentum, which
energy-momentum turns out to be just that proposed by Dougan and Mason [5]
within the GHP formalism. This justifies the rather heuristic
argumentation-based choice of Dougan and Mason for the spinor propagation
laws. The part of section 2 dealing with the holomorphy/anti-holomorphy
operators can therefore be considered as the investigation of the
conditions under which the Dougan--Mason construction can(not) be done.
Although there are `exceptional' 2-surfaces, e.g. the marginally trapped
surfaces, for which the Dougan--Mason construction does not work (at least
in its present form), for `strictly convex' 2-spheres the construction
seems to be well defined.\par
       In section 3 first the Dougan--Mason energy and mass nonnegativity
proof will be reviewed in the covariant spinor formalism and then
some recent results on the Dougan--Mason energy-mom\-en\-tum will be
generalized and discussed from new points of view too. We give equivalent
statements for the vanishing of the quasi-local energy-momentum and for the
vanishing of the quasi-local mass. In particular we will see that the
vanishing of the Dougan--Mason mass (energy-momentum) is equivalent to the
existence of one (two) Sen--constant spinor field(s) on the 2-surface $\$
$. Furthermore, it will be clear that the {\it pp}-wave Cauchy
developments with pure radiation can be characterized not only by the
usual Cauchy data on a finite {\it three dimensional} Cauchy surface
$\Sigma$, but by the {\it two dimensional} Sen--geometry of the boundary
of $\Sigma$ too. \par
        Finally we examine the possibility of defining the quasi-local mass
of marginally trapped surfaces as the limit of the masses of a family of
non-exceptional 2-spheres. We will see that this definition is ambiguous
since the limit depends not only on the geometry of the trapped surface
but the family as well. \par
       The notations and conventions are the same that used in [1]. In
particular, abstract index formalism [8] will be used unless otherwise
stated. The `name' or component indices will be underlined. \par
\vskip 1truecm

{\lbf 1. Quasi-local Nester--Witten integrals}\par
\vskip 0.5truecm

\noindent First recall that for any two spinor fields $\lambda_A$, $\mu_A$
the spinorial Nester--Witten 2-form is defined [12,13] by:

$$
u(\lambda,\bar\mu)_{ab}:={i\over2}\Bigl(\bar\mu_{A^\prime}\nabla_{BB^\prime
}\lambda_A-\bar\mu_{B^\prime}\nabla_{AA^\prime}\lambda_B\Bigr).\eqno(1.1)
$$

\noindent Apart from an exact form this is `hermitian' in the sense that

$$
u(\lambda,\bar\mu)_{ab}={\overline{u(\mu,\bar\lambda)}}_{ab}-{i\over2}
\bigl(\nabla_aK_b-\nabla_bK_a\bigr), \hskip 1truecm
K_a:=\lambda_A\bar\mu_{A^\prime}.\eqno(1.2)
$$

\noindent Thus its real and imaginary parts are $F_{ab}:=u(\lambda,\bar
\mu)_{ab}+{\overline{u(\lambda,\bar\mu)}}_{ab}=$ $u(\lambda,\bar\mu)_{ab}
+u(\mu,\bar\lambda)_{ab}$ $+{i\over2}(\nabla_a\bar K_b-\nabla_b\bar K_a)$ and
$iK_{ab}:=u(\lambda,\bar\mu)_{ab}-{\overline{u(\lambda,\bar\mu)}}_{ab}$
$=u(\lambda,\bar\mu)-u(\mu,\bar\lambda)_{ab}$ $-{i\over2}(\nabla_a\bar K_b
-\nabla_b\bar K_a)$, respectively. For $\mu_A=\lambda_A$  $F_{ab}$ is just
the Reula--Tod 2-form [14] (and its dual $*F_{ab}$ is the Ludvigsen--Vickers
2-form [15]) by means of which they proved the positivity of the
Bondi--Sachs mass. In general $K_{ab}$ is not exact. In the non-abstract
index formalism Sparling's form is

$$
\Gamma(\lambda,\bar\mu):=i\nabla_{BB^\prime}\lambda_A\nabla_{CC^\prime}
\bar\mu_{A^\prime}dx^a\wedge dx^b\wedge dx^c.\eqno(1.3)
$$

\noindent This is `hermitian': ${\overline{\Gamma(\lambda,\bar\mu)}}=\Gamma
(\mu,\bar\lambda)$, and the Sparling equation is

$$
du=\Gamma-{1\over2}\lambda^A\bar\mu^{A^\prime}G_{AA^\prime}{}^b\Sigma_b,
\eqno(1.4)
$$

\noindent where $G_{ab}$ is the Einstein tensor, $\Sigma_a:={1\over3!}
\varepsilon_{abcd}dx^b\wedge dx^c\wedge dx^d$ and Einstein's equations are
written in the form $G_{ab}=-\kappa T_{ab}$. \par
      Let us define the quasi-local Nester--Witten integral by

$$
H_{\$}[\lambda_R,\bar\mu_{S^\prime}]:={2\over\kappa}\oint_{\$}u(\lambda,\bar
\mu)_{ab}dx^a\wedge dx^b={4\over\kappa}\oint_{\$}t^av^b*u_{ab}d\$,\eqno(1.5)
$$

\noindent which in the formalism developed in [1] takes the following form

$$
H_{\$}[\lambda_R,\bar\mu_{S^\prime}]={2\over\kappa}\oint_{\$}\bar
\gamma^{R^\prime S^\prime}\bar\mu_{R^\prime}\Delta_{S^\prime}{}^S\lambda_S
d\$={2\over\kappa}\oint_{\$}\bar\mu^{R^\prime}\bigl(\Delta_{R^\prime}^{+R}
\lambda_R-\Delta_{R^\prime}^{-R}\lambda_R\bigr)d\$.\eqno(1.6)
$$

\noindent By virtue of (1.2) $H_{\$}$ is a hermitian bilinear functional
on the space $C^\infty(\$,{\bf S}_A)$ of smooth spinor fields on $\$ $.
The importance of the quasi-local Nester--Witten integral is shown by
Sparling's equation (1.4): $H_{\$}$ is connected to the energy-momentum of
gravitating systems [12-15]. It might be interesting to note that the
integral of the conformal invariant hermitian scalar product of local
twistors on $\$ $ can also be expressed by (1.6): If ${\tt Z}^\alpha=(
\lambda^A,i\Delta_{A^\prime K}\lambda^K)$ and are ${\tt W}^\beta=(\mu^B,
i\Delta_{B^\prime K}\mu^K)$ are local twistors on $\$ $ [1] and $\langle{
\tt Z},\bar{\tt W}\rangle:=-\lambda^Ai\Delta_{AA^\prime}\bar\mu^{A^\prime}
+\bar\mu^{A^\prime}i\Delta_{AA^\prime}\lambda^A$ then

$$
h_{\$}[\lambda_R,\bar\mu_{S^\prime}]:=H_{\$}[\gamma_R{}^K\lambda_K,\bar\mu
_{S^\prime}] -H_{\$}[\lambda_R,\bar\gamma_{S^\prime}{}^{K^\prime}\bar
\mu_{K^\prime}]={2i\over \kappa}\oint_{\$}\langle{\tt Z},\bar{\tt W}\rangle
d\$.\eqno(1.7)
$$

\noindent If in (1.6) the spinor field $\mu_S$ is chosen to be $\Delta_S
{}^{S^\prime}\bar\omega_{S^\prime}$ for some spinor field $\omega_S$ then
by the first Sen--Witten type identy (7.3) of [1]

$$
\eqalign{A_{\$}[\lambda_R,\omega_S]:&=H_{\$}[\lambda_R,\Delta_{S^\prime}
{}^S\omega_S]=\cr
&={2\over\kappa}\oint_{\$}\Bigl(\bigl(\Delta^{+A}_{R^\prime}\omega_A\bigr)
\bigl(\Delta^{-R^\prime B}\lambda_B\bigr)+\bigl(\Delta^{+A}_{R^\prime}
\lambda_A\bigr)\bigr(\Delta^{-R^\prime B}\omega_B\bigr)\Bigr)d\$=\cr
&={2\over\kappa}\oint_{\$}\Bigl(\bigl({\cal T}^+_{R^\prime RS}{}^A\omega_A
\bigr)\bigl({\cal T}^{-R^\prime RSB}\lambda_B\bigr)+\bigl({\cal T}^+_{
R^\prime RS}{}^A\lambda_A\bigr)\bigl({\cal T}^{-R^\prime RSB}\omega_B
\bigr)\Bigr)d\$-\cr
&-{i\over\kappa}\oint_{\$}\lambda^A\omega^BR_{ABcd}dx^c\wedge dx^d.\cr}
\eqno(1.8)
$$

\noindent This is a symmetric bilinear functional on $C^\infty(\$,
{\bf S}_A)$. If at least one of $\lambda_R$ and $\omega_S$ is a solution
of the 2-surface twistor equation, or both $\lambda_R$, $\omega_S$ belong
to the kernel space of ${\cal T}^+$ or to the kernel space of ${\cal T}^-$
then $A_{\$}$ reduces to Penrose's charge integral; i.e. to the expression
of the kinematic twistor [2,12,16]. Or, in other words, for spinor fields
satisfying the twistor equation Penrose's charge integral can be expressed
as a quasi-local Nester--Witten integral too (see also [13]). In a similar
way one can choose $\mu_S:=\gamma_S{}^K\Delta_K{}^{K^\prime}\bar\omega
_{K^\prime}$ for some spinor field $\omega_S$ in (1.6). Then

$$
\eqalign{I_{\$}[\lambda_R,\omega_S]:&=H_{\$}[\lambda_R,\bar\gamma_{
S^\prime}{}^{K^\prime}\Delta_{K^\prime}{}^K\omega_K]=\cr
&={2\over\kappa}\oint_{\$}\Bigl(\bigl(\Delta^{+A}_{R^\prime}\omega_A\bigr)
\bigl(\Delta^{-R^\prime B}\lambda_B\bigr)-\bigl(\Delta^{+A}_{R^\prime}
\lambda_A\bigr)\bigl(\Delta^{-R^\prime B}\omega_B\bigr)\Bigr)d\$}\eqno(1.9)
$$

\noindent is an antisymmetric bilinear functional on $C^\infty(\$,{\bf
S}_A)$ and by the second Sen--Witten type identity (7.4) of [1] $I_{\$}[
\lambda_R,\omega_S]$ can be rewritten as the integral of a quadratic
expression of the chiral twistor-derivatives and the charge integrals of
the curvature and the torsion. The kernel space of these functionals is

$$
\eqalign{\ker H_{\$}:&=\bigl\{\lambda_A\in C^\infty(\$,{\bf S}_A)\vert
\hskip 3pt H_{\$}[\lambda,\bar\mu]=0 \hskip 5pt \forall\mu_A\in C^\infty(\$,
{\bf S}_A)\bigr\}=\cr
&=\ker A_{\$}=\ker I_{\$}=\ker\bigl(\Delta^-\oplus\Delta^+\bigr).\cr}
\eqno(1.10)$$

\noindent $\ker\Delta^\pm$ are infinite dimensional subspaces of $C^\infty
(\$,{\bf S}_A)$.\par
      In the present paper we are interested in the possibility of finding
quasi-local energy-momentum (and possibly angular momentum) expressions for
gravitating systems in the form of Nester--Witten integrals. However these
quantities are expected to be in the dual space of the four real dimensional
vector space of the `quasi-translations' and of the six real dimensional
vector space of the `quasi-rotations' of $\$ $, respectively. Furthermore in
order to define the quasi-local mass as the length of the quasi-local
energy-momentum the space of `quasi-translations' must have a Lorentzian
metric. Thus what we need is a rule to reduce the infinite dimensional
complex vector space $C^\infty(\$,{\bf S}_A)$ to a finite dimensional
subspace that can be interpreted as the space of the spinor
constituents of the `quasi-translations'/`quasi-rotations' of $\$ $. In
other words, propagation law(s) for the spinor fields $\lambda_R$ and
$\mu_S$ should be prescribed. It is natural to look for these propagation
laws in the form $\D\lambda=0$ where $\D$ is a {\it differential} operator
acting on the space of the spinor fields. Since the `quasi-translations'
(the `quasi-rotations') should form finite dimensional vector space(s) $\D$
must be a {\it linear} differential operator with {\it finite dimensional
kernel}. \par
     Since $\ker H_{\$}=\ker A_{\$}=\ker I_{\$}=\ker (\Delta^-\oplus
\Delta^+)$ one can see that the full 2 dimensional Weyl--Sen--Witten
equation, in contrast to the 3 dimensional Sen--Witten equation,
cannot be used to define propagation law(s) within $\$ $. Thus to find the
appropriate propagation law(s) a detailed and systematic study of the linear
differential operators on $C^\infty(\$,{\bf S}_A)$ is needed.\par
\vskip 1truecm

{\lbf 2. Propagation laws}\par \vskip 0.5truecm

\noindent In this paper we restrict our considerations to those {\it first
order} linear differential operators $\D$ on $C^\infty(\$,{\bf S}_A)$ that
can be constructed {\it only from} the chiral irreducible parts of the Sen
operator $\Delta_e$. (See the Introduction.) In the GHP formalism that we
will use in this section are the chiral irreducible parts $\Delta^\pm$ and
${\cal T}^\pm$ of $\Delta_e$. They are differential operators on $E^\infty
(-1,0)\oplus E^\infty(1,0)$ $\simeq C^\infty(\$,{\bf S}_A)$:

$$\eqalign{
-\Delta^-\lambda=\edth\lambda^0-\rho^\prime\lambda^1,\hskip 2truecm
&-\Delta^+\lambda=\edth^\prime\lambda^1-\rho\lambda^0\cr
-{\cal T}^-\lambda=\edth\lambda^1-\sigma\lambda^0,\hskip 2truecm
&-{\cal T}^+\lambda=\edth^\prime\lambda^0-\sigma^\prime\lambda^1,\cr}
\eqno(2.1)$$

\noindent where the spinor components are defined by $\lambda^R=:\lambda^0
o^R+\lambda^1\iota^R$ $=:\lambda_1o^R-\lambda_0\iota^R$ (see (6.9-12) of
[1]). These are the `elementary operators' by means of which we construct
all the first order operators. Higher order operators can also be
constructed from $\Delta^\pm$ and ${\cal T}^\pm$ taking into account that
by (6.13) of [1] they cannot be composed in any way. The properties of the
edth operators we need are clarified in the Appendix. \par
       The irreducible chiral operators $\Delta^\pm$ and ${\cal T}^\pm$
have infinite dimensional kernel spaces, and the direct sums $\Delta^\pm
\oplus\Delta^\pm$ and ${\cal T}^\pm\oplus{\cal T}^\pm$ are obviously
equivalent to $\Delta^\pm$ and ${\cal T}^\pm$ themselves, respectively.
The remaining direct sums consisting of two terms are the operators

$$
\eqalign{-\Delta :=\Delta^-\oplus\Delta^+:&E^\infty(-1,0)\oplus
   E^\infty(1,0)\rightarrow E^\infty (0,-1)\oplus E^\infty(0,1)\cr
-{\cal H}^-:=\Delta ^-\oplus {\cal T}^-:&E^\infty(-1,0)\oplus
   E^\infty(1,0)\rightarrow E^\infty(0,-1)\oplus E^\infty(2,-1)\cr
-{\cal C}^-:={\cal T}^+\oplus\Delta^-:&E^\infty(-1,0)\oplus
   E^\infty(1,0)\rightarrow E^\infty(-2,1)\oplus E^\infty(0,-1)\cr
-{\cal C}^+:= \Delta^+\oplus{\cal T}^-:&E^\infty(-1,0)\oplus
   E^\infty(1,0)\rightarrow E^\infty(0,1)\oplus E^\infty(2,-1)\cr
-{\cal H}^+:={\cal T}^+\oplus\Delta^+:&E^\infty(-1,0)\oplus
   E^\infty(1,0)\rightarrow E^\infty(-2,1)\oplus E^\infty(0,1)\cr
-{\cal T}:={\cal T}^+\oplus{\cal T}^-:&E^\infty(-1,0)\oplus
   E^\infty(1,0)\rightarrow E^\infty(-2,1)\oplus E^\infty(2,-1).\cr}
\eqno(2.2)$$

\noindent $\Delta$, ${\cal T}$ and ${\cal H}^\pm$ are elliptic operators,
and since $\$ $ is compact they have finite dimensional kernels. To
determine the dimension of the kernels first calculate their analitic
index following Baston's calculations [10] for the twistor operator. Recall
that the index of an elliptic operator $\D$ is $\dim\ker\D-\dim\ker\D
^\dagger$, where $\D^\dagger$ is the adjoint of $\D$ with respect to some
hermitian scalar products on the space of the smooth sections of the vector
bundles. By the Atiyah--Singer index theorem the index of the elliptic
linear differential operators is a topological invariant of the bundles
and the operators. Thus their index coincides with that of the operators

$$
\left(\matrix{\edth_{(-1,0)}&0\cr 0&\edth^\prime_{(1,0)}\cr}\right),
\left(\matrix{\edth^\prime_{(-1,0)}&0\cr 0&\edth_{(1,0)}\cr}\right),
\left(\matrix{\edth^\prime_{(-1,0)}&0\cr0&\edth^\prime_{(1,0)}\cr}\right),
\left(\matrix{\edth_{(-1,0)}&0\cr0&\edth_{(1,0)}\cr}
\right),\eqno(2.3)$$

\noindent respectively. Then the index of the elliptic operators can be
calculated from eq.(A.1) of the Appendix. If ${\cal G}$ is the genus of
$\$ $ then

$$\eqalign{ \index \Delta &=0,\cr
\index {\cal T}&=4(1-{\cal G} ), \cr
\index {\cal H}^{\pm }&=2(1-{\cal G}).\cr}
\eqno(2.4)$$

\noindent
Thus from the index theorem it does not follow the existence of non
trivial solutions of $\Delta_{R^\prime}{}^R\lambda_R=0$. Since however
$\ker\Delta$ is precisely the kernel of $H_{\$}$, $A_{\$}$ and $I_{\$}$,
$\Delta_{R^\prime}{}^R$ could not be used to define propagation laws even
if there were nontrivial solutions of $\Delta_{R^\prime}{}^R\lambda_R=0$.
\par
       By (2.4) for topological 2-spheres ${\cal T}_{R^\prime RS}{}^K
\lambda_K=0$ has {\it at least} four independent solutions [10]. One way of
determining $\dim\ker{\cal T}$ is to consider the adjoint ${\cal T}
^\dagger$ of the twistor operator. For any fixed nowhere vanishing
$h\in E^\infty(1,1)$ $\langle(\phi^0,\phi^1),(\psi^0,\psi^1)\rangle$ $:=
\langle\phi^0,\psi^0\rangle_{(p,q)}$ $+\langle\phi^1,\psi^1\rangle_{(
p^\prime,q^\prime)}$ is a hermitian scalar product on the space of the
smooth sections of the Whitney sum $E(p,q)\oplus E(p^\prime,q^\prime)$,
where $\langle\phi,\psi\rangle$ is defined by (A.3) in the
Appendix. With respect to this scalar product the adjoint of the twistor
operator is

$$\eqalign{
{\cal T}^\dagger:&E^\infty(-2,1)\oplus E^\infty(2,-1)\rightarrow E^\infty
(-1,0)\oplus E^\infty(1,0)\cr
&(\omega^0,\omega^1)\mapsto \Bigl(-{1\over{\mid h\mid}}\bigl(\edth\bigl(
\mid h\mid\omega^0\bigr)+\bar\sigma{\omega^1\over{\mid h\mid}}\bigr),
-\mid h\mid\bigl(\edth^\prime\bigl({\omega^1\over{\mid h\mid}}\bigr)+
\bar\sigma^\prime\mid h\mid\omega^0\bigr)\Bigr).\cr}\eqno(2.5)
$$

\noindent Thus with the definitions $\mu^0:=\mid h\mid\omega^0\in E^\infty
(-1,2)$, $\mu^1:=\mid h\mid^{-1}\omega^1\in E^\infty(1,-2)$ the adjoint
of the twistor equations are read

$$
\eqalign{\edth_{(-1,2)}\mu^0+\bar\sigma\mu^1&=0\cr
\edth^\prime_{(1,-2)}\mu^1+\bar\sigma^\prime\mu^0&=0.\cr}\eqno(2.6)
$$

\noindent Since by (A.9) $\dim\ker\edth_{(-1,2)}=\dim\ker\edth^\prime
_{(1,-2)}=0$, $\dim\ker{\cal T}^\dagger=0$ if $\sigma=0$ or $\sigma^\prime
=0$ on $\$ $, and hence in these cases $\ker{\cal T}$ is precisely four
dimensional. In particular, for round spheres (i.e. for 2-spheres of
spherical symmetry in spherically symmetric spacetimes [17]) the number of
independent solutions of ${\cal T}_{R^\prime RS}{}^K\lambda_K=0$ is in
fact precisely four [12]. However, as Jeffryes [18] has shown, in general
$\dim\ker{\cal T}$ may be greater than $\index {\cal T}$. Thus for the
choice $\D={\cal T}$ we have {\it at least} 4 real and 6 complex
integrals $H_{\$}$, 10 complex integrals $A_{\$}$
and 6 complex integrals $I_{\$}$. Since we would like to have four real
integrals for the energy-momentum and six real integrals for the angular
momentum none of the expressions (1.6), (1.8) and (1.9) seems to yield
the expected number of kinematical quantities unless an extra structure
is used to reduce the number of them. In fact, in certain special cases
(1.9) defines a real, skew and simple twistor ${\tt I}_{\alpha\beta}$,
the so-called infinity twistor, so that ${\tt I}_{\alpha\beta}$, together
with the hermitian metric ${\tt h}_{\alpha\beta^\prime}$, defined by (1.7),
can be used to reduce the ten complex components of the kinematic twistor
${\tt A}_{\alpha\beta}$, defined by (1.9), to ten real components [2,12,16].
This is the original twistor-theory-motivated proposal of Penrose for the
four quasi-local energy-momentum and six angular momentum. In general,
however, no such infinity twistor exists [19].\par
     Since $\$ $ is an oriented closed 2 dimensional Rimannian submanifold
it is a compact Riemann surface with the naturally defined complex
structure [9,20-22]. This is precisely the integrable almost complex
structure that the projections $\pi^{\pm a}{}_b:=\pi^{\pm A}{}_B\bar\pi^{
\mp A^\prime}{}_{B^\prime}$ define (see [1]). One can therefore define
holomorphic functions, the multiplicity of the zeros and the order of the
poles of meromorphic functions. If $(U,\xi)$ is a local holomorphic
coordinate system and $m^a=P(\xi,\bar\xi)({\partial\over\partial\xi})^a$
for some nonzero smooth $P(\xi,\bar\xi)$ then $f:U\rightarrow{\bf C}$ is
holomorphic iff $\bar m^a\delta_a f=0$ iff $\bar m^a\Delta_a f=0$. Thus
the notion of holomorphic (meromorphic) functions and the multiplicity of
their zeros and the order of their poles are independent of the operators
$\delta_a$, $\Delta_a$. The notion of the holomorphic tensor and spinor
fields, however, does depend on the choice of the differential operator.
The spinor field $\phi^{A...A^\prime...}_{B...B^\prime...}$ is said to
be holomorphic {\it with respect to the induced Levi--Civit\`a connection}
if $\bar m^e\delta_e\phi^{A...A^\prime...}_{B...B^\prime...}=0$. In the
case of {\it surface} vector fields this notion of holomorphy coincides
with the holomorphy with respect to the {\it intrinsic} complex structure
of $\$ $ (see the Appendix). The point $p\in\$ $ is said to be a zero of
the $\delta_a$-holomorphic spinor field $\phi^{...}{}_{...}$ with
multiplicity $m$ if $\phi^{...}{}_{...}$ and its first $(m-1)$ $m^a
\delta_a$-derivatives vanish at $p$ but its $m$th $m^a\delta_a
$-derivative is not zero there. It is not difficult to show that $p$ is a
zero of the $\delta_a$-holomorphic $\phi^{...}{}_{...}$ with multiplicity
$m$ if and only if there is a $\delta_a$-holomorphic spinor field $\psi
^{...}{}_{...}$ and a holomorphic function $f$ on an open neighbourhood
$W$ of $p$ such that $\phi^{...}{}_{...}=f \psi^{...}{}_{...}$ and $\psi
^{...}{}_{...}$ is nonzero on $W$ and $p$ is a zero of $f$ with
multiplicity $m$. (Hint: On a sufficiently small neighbourhood $W$ of $p$
there are $\delta_a$--holomorphic and $\delta_a$--anti-holomorphic spinor
fields $\lambda^{\uA}_R$ and $\mu^{\uB}_R$, ${\uA},{\uB}=0,1$,
respectively, which form bases
in the spinor space at each point of $W$. Then the components $\phi_{\uA
...\uB {\uC}^\prime...{\uD}^\prime}$ of $\phi_{A...BC^\prime...D^\prime}$
in the $\delta_a$--holomorphic basis $\lambda^{\uA}_A ...\lambda^{\uB}_B
\bar\mu^{{\uC}^\prime}_{C^\prime}...\bar\mu^{{\uD}^\prime}_{D^\prime}$ are
holomorphic.) As a consequence the zeros of a not identically vanishing
$\delta_a$--holomorphic spinor field are isolated. The spinor field
$\phi^{...}{}_{...}$ is said to be holomorphic {\it with respect to the 2
dimensional Sen connection} if ${\bar m}^e\Delta_e\phi^{...}{}_{...}=0$.
Thus the spinor field $\lambda_R$ is holomorphic with respect to $\Delta_a$
iff $\Delta^+_{R^\prime}{}^R\lambda_R=0$ {\it and} ${\cal T}^+_{R^\prime RS}
{}^K\lambda_K=0$ by (6.10) and (6.12) of [1]; and $\lambda_R$ is
anti-holomorphic with respect to $\Delta_a$ iff $\Delta_{R^\prime}^{-R}
\lambda_R=0$ {\it and} ${\cal T}^-_{R^\prime RS}{}^K\lambda_K=0$ by (6.9)
and (6.11) of [1]. Thus ${\cal H}^\pm$ may be called the
holomorphy/anti-holomorphy operators. One can
define the multiplicity of the zeros of $\Delta_a$-holomorphic spinor
fields too, and their zeros are also isolated. The notion of holomorphic
spinor fields defined by Dougan and Mason [5] coincides with this second
notion of holomorphy, and, apart from the Appendix, in the rest of this
paper `holomorphy' will mean `holomorphy with respect to $\Delta_a$' unless
otherwise stated. \par
       Since ${\cal H}^-$ is elliptic and $\$ $ is compact, $\ker{\cal H}^-$
is finite dimensional. By (2.4) for topological 2-spheres $\index{\cal H}^-
=2$; i.e. there are {\it at least} two linearly independent anti-holomorphic
spinor fields, say $\lambda^0_R$ and $\lambda^1_R$, on $\$ $. (If a spinor
field has a `name' index too then the spinor index will be written as a
subscript and the `name' index as a superscript. Thus for example $\lambda^0
_1$ is the {\it 1 component} of the {\it zeroth spinor}: $\lambda^0_1=
\lambda^0_R\iota^R$, while $\lambda^1_0$ is the {\it 0 component} of
the {\it first spinor}: $\lambda^1_0=\lambda^1_Ro^R$.) Then
$m^e\nabla_e(\varepsilon^{RS}\lambda^{\uA}_R\lambda^{\uB}_S)=0$,
${\uA},{\uB}=0,1$; and hence by Liouville's theorem $\varepsilon^{\uA
\uB}:=\varepsilon^{RS}\lambda^{\uA}_R\lambda^{\uB}_S$ is constant on $\$ $.
If this constant is not zero (such two-surfaces will be called generic)
then it can be chosen to be $\epsilon^{{\uA}{\uB}}$, the Levi--Civit\`a
alternating symbol, and hence $\lambda^0_R$ and $\lambda^1_R$ form a
normalized basis in the spinor space at each point of $\$ $. Thus for any
spinor field $\lambda_R$ there are complex functions $\alpha$ and $\beta$
on $\$ $ such that $\lambda_R=\alpha\lambda^0_R+\beta\lambda^1_R$. If
$\lambda_R$ is
anti-holomorphic then $\alpha$ and $\beta$ are anti-holomorphic on $\$ $.
Then by Liouville's theorem they must be constant; i.e. for generic
2-spheres $\dim_{\bf C}\ker{\cal H}^-=2$. If $\lambda^0_R\lambda^1_S
\varepsilon^{RS}=0$ (such two-surfaces will be called exceptional) then
there must be a function $f:\$\rightarrow{\bf C}$ such that $m^a\Delta_a
f=0$ and $\lambda^1_R=f\lambda^0_R$. But $\lambda^0_R$ and $\lambda^1_R$
can be independent solutions to ${\cal H}^-\lambda=0$ only if $\lambda^0_R$
has at least one zero and $f$ is only anti-meromorphic with pole(s) at the
zero(s) of $\lambda^0_R$. Since the sum of the order of the poles of an
anti-meromorphic function on a sphere is equal to the sum of the
multiplicity of its zeros, the sum of the multiplicity of the zeros of
$\lambda^0_R$ is equal to the sum of the multiplicity of the zeros of
$\lambda^1_R$. If $\lambda_R$ is any anti-holomorphic spinor field on
$\$ $ then $\lambda_R\lambda^0_S\varepsilon^{RS}=0$ since $\lambda_R
\lambda^0_S\varepsilon^{RS}$ is constant on $\$ $ and $\lambda^0_R$ has
zeros. Thus for exceptional 2-spheres any anti-holomorphic spinor field
on $\$ $ is proportional to a `basis solution', say $\lambda^0_R$, and the
factor of proportionality is an anti-meromorphic function. These functions
can be given explicitly in a coordinate system: Let $n\in\$ $ (`north
pole'), $U:=\$-\{n\}$ and $\xi:U\rightarrow {\bf C}$ a holomorphic
coordinate. Let the zeros of $\lambda^0_R$ be $z_0=\infty$ and $z_1,
...,z_k\in\xi(U)={\bf C}$ with multiplicities $m_0,m_1,...,m_k$,
respectively, and define $m:=m_0+m_1+...+m_k$. Then the most general
anti-meromorphic function on $\$ $ whose possible poles are $z_0,...,
z_k$ with maximal order $m_0,...,m_k$, respectively, is

$$
f(\bar\xi)={a_m\bar\xi^m +a_{m-1}\bar\xi^{m-1}+ ...+a_1\bar\xi+a_0 \over
(\bar \xi -z_1)^{m_1} ...(\bar\xi -z_k)^{m_k} },\eqno(2.7)
$$

\noindent where $a_0,a_1,...,a_k\in{\bf C}$. These functions
form a complex vector space of dimension $m+1$, and hence $\dim\ker{\cal
H}^-=m+1$. To decide whether there exist exceptional 2-spheres first
consider the adjoint of the equation ${\cal H}^-\lambda=0$. This reads

$$
\eqalign{\edth^\prime\mu^0+\bar\sigma\mu^1&=0\cr
\edth^\prime\mu^1+\rho^\prime\mu^0&=0,\cr}\eqno(2.8)
$$

\noindent where $\mu^0\in E^\infty(1,0)$ and $\mu^1\in E^\infty(1,-2)$.
By (A.9) $\dim\ker\edth^\prime_{(1,0)}=\dim\ker\edth^\prime_{(1,-2)}=0$,
thus the adjoint equation (2.8) has only the trivial solution if
$\rho^\prime=0$ or $\sigma=0$ on the whole $\$ $, and $\ker{\cal H}^-$ is
still two dimensional. First consider the special case $\rho^\prime=0$,
which turns out to characterize an exceptional 2-sphere.\par

\medskip \noindent
{\vbox {\bf Lemma 2.9:} If $\lambda_R\in\ker{\cal H}^-$ then the following
statements are equivalent:
\item{i.} $\lambda_R=\lambda^1 \iota_R$, i.e. $\lambda^0$ is zero on $\$ $,
\item{ii.} $\$ $ is a future marginally trapped surface, i.e.
          $\rho^\prime=0$ on $\$ $,
\item{iii.} $\lambda_R$ is anti-holomorphic with respect to $\delta_e$ too.
\par
\noindent If any of these conditions is satisfied then $\dim\ker{\cal H}^-
=2$, the two independent anti-holomorphic spinor fields are proportional,
$\lambda^1_R=f\lambda^0_R$, and each has a single zero with multiplicity
1.} \par
\bigskip

\noindent This lemma is a simple consequence of (A.9), (A.11), (A.18)
and the GHP form of (A.17).\par
      Now consider a smooth 1 parameter family $\$(u)$ of spacelike
2-spheres, $u\in(-\epsilon,\epsilon)$, such that $\$(0)=\$ $ is a
future marginally trapped surface. We show that for sufficiently small
$\mid u\mid\not=0$ $\$(u)$ is not exceptional provided $\dot\rho^\prime:=
({d\over du}\rho^\prime(u))_{u=0}$ is nowhere zero on $\$ $. If ${\cal
H}^-_u$ is the anti-holomorphy operator for $\$(u)$ then in the complex
coordinate system $(\xi,\bar\xi)$ above the equation ${\cal H}^-_u
\lambda(u)=0$ takes the following form

$$\eqalign{
P(u){\partial\lambda_1(u)\over\partial\xi}+\beta(u)\lambda_1(u)
+\rho^\prime(u)\lambda_0(u)&=0\cr
P(u){\partial\lambda_0(u)\over\partial\xi}-\beta(u)\lambda_0(u)+\sigma(u)
\lambda_1(u)&=0.\cr}\eqno(2.10)
$$

\noindent Here $\beta$ is the spin coeffitient $-B_em^e=m^e\nabla_e
\iota_Ao^A$ and $\lambda_R(u)=:\lambda_1(u)o_R-\lambda_0(u)\iota_R$. Since
$\index{\cal H}_u^-=2$, there certainly exist two linearly independent
solutions $\lambda^{\uA}_R(u)$, ${\uA}=0,1$, to ${\cal H}^-_u\lambda
(u)=0$ for any $u\in(-\epsilon,\epsilon)$; and these solutions can be
chosen so that $\lambda^{\uA}_R(0)$ to coincide with the independent
solutions (A.18) guaranteed by Lemma 2.9. Then taking the derivative of
(2.19) with respect to $u$ at $u=0$ we have

$$\eqalignno{
\edth\dot\lambda^{\uA}_1 +\dot\rho^\prime\lambda^{\uA}_0&=0&(2.11)\cr
\edth\dot\lambda^{\uA}_0 +\dot{\edth}\lambda^{\uA}_0 +\sigma\dot\lambda
^{\uA}_1&=0,&(2.12)\cr}
$$

\noindent where $\dot{\edth}\lambda^{\uA}_0:=\dot P{\partial\over\partial
\xi}\lambda^{\uA}_0-\dot\beta\lambda^{\uA}_0$. $\dot\lambda^{\uA}_1$
therefore depend only on $\lambda^{\uA}_0$ and $\dot\rho^\prime$, but are
independent all of $\dot P$, $\dot\beta$ and $\dot\sigma$. Since
$\varepsilon^{RS}\lambda^0_R(u)\lambda^1_S(u)$ is constant on $\$(u)$, its
derivative is also constant on $\$ $, and since

$$
{d\over du}\Bigl(\varepsilon^{RS}\lambda^0_R(u)\lambda^1_S(u)\Bigr)_{u=0}
=\lambda^0_0\bigl(\dot\lambda^1_1-f\dot\lambda^0_1\bigr),\eqno(2.13)
$$

\noindent it is zero if and only if $\dot\lambda^1_1=f\dot\lambda^0_1$.
Since by (A.9) $\dim\ker\edth_{(-1,0)}=0$, $\dot\lambda^1_1=f
\dot\lambda^0_1$ iff $f\dot\lambda^0_1$ is smooth; i.e. iff
$\dot\lambda^0_1$ has a zero precisely at the pole of $f$. In other
words the derivative (2.13) is zero if and only if the solution
$\dot\lambda^0_1$ of (2.11) has a zero and the zero of $\dot\lambda^0_1$
and $\lambda^0_0$ coincide. We will show that the zero of $\dot\lambda^0_1$
and $\lambda^0_0$ do not coincide provided $\dot\rho^\prime\not=0$
everywhere on $\$ $; and hence for sufficiently small nonzero $\mid u\mid$
the 2-surface $\$(u)$ is generic. The coordinate system $(\xi,\bar\xi)$
can always be chosen so that $P(\xi,\bar\xi;u)=e^{-\omega(u)}(1+\xi
\bar\xi)$, where $\omega(u)$ is a smooth real function on the whole $\$ $;
and by (A.18) $\lambda^0_0=i({\nu\over\bar P})^{1\over 2}$ where $\nu=\exp(
-\int_0^\xi A_e)$. With these choices the solution of (2.11) is

$$
\dot\lambda^0_1(\xi,\bar\xi)=-ie^{-{1\over2}\omega+{1\over2}\int A_e}
\sqrt{1+\xi\bar\xi}\int_0^\xi {\dot\rho^\prime(\xi^\prime,\bar\xi)\over
(1+\xi^\prime\bar\xi)^2}e^{2\omega-\int A_e}d\xi^\prime, \eqno(2.14)
$$

\noindent and hence

$$
\mid\dot\lambda^0_1(\xi,\bar\xi)\mid\ge\min_{\$}\bigl\{\mid e^{-{1\over2}
\omega+{1\over2}\int A_e}\mid\bigr\} \min_{\$}\bigl\{\mid\dot\rho^\prime
e^{2\omega-\int A_e}\mid\bigr\}{\mid \xi\mid\over\sqrt{1+\xi\bar\xi}}.
\eqno(2.15)$$

\noindent Thus $\dot\lambda^0_1(\xi,\bar\xi)$ may have a zero only in the
south pole ($\xi=0$) while $\lambda^0_0$ has a single zero in the
north pole ($\xi=\infty$). This result can also be interpreted as the
perturbations of a future marginally trapped surface yield generic
2-spheres provided the perturbations satisfy $\dot\rho^\prime\not=0$
everywhere. The marginally trapped surfaces are therefore really
`exceptional'.\par
       Next suppose that $\sigma=0$. Then for the two linearly independent
solutions, $\lambda^0_R$ and $\lambda^1_R$, the equation ${\cal H}^-\lambda
=0$ reduces to

$$\eqalignno{\edth\lambda^{\uA}_1+\rho^\prime\lambda^{\uA}_0&=0&(2.16a)\cr
\edth\lambda^{\uA}_0&=0.&(2.16b)\cr}
$$

\noindent But (2.16b) is just the GHP form of (A.17), furthermore (2.16a)
has the same structure as (2.11). Thus the pair $(\lambda^0_0,\lambda^0_1)$
is a solution of (2.16) where in the complex coordinate system above
$\lambda^0_0=i(1+\xi\bar\xi)^{-{1\over2}}e^{{1\over2}\omega-{1\over2}\int
A}$ and $\lambda^0_1$ is given by (2.14) (without the dots). $\lambda^0_0$
has a zero in the north pole ($\xi=\infty$), but in general (e.g. if
$\rho^\prime\ge 0$ but not identically zero on $\$ $; i.e. $\$ $ is
`convex') the solution $(\lambda^0_0,\lambda^0_1)$ does not have a zero.
However with an appropriately chosen $\rho^\prime$ $\lambda^0_1$ will have
a zero in the north pole too. For example with $\rho^\prime={1-\xi\bar\xi
\over1+\xi\bar\xi}$ we have

$$\eqalign{
\mid\lambda^0_1(\xi,\bar\xi)\mid&\leq\max_{\$}\bigl\{\mid e^{-{1\over2}
\omega+{1\over2}\int A_e}\mid\bigr\}\max_{\$}\bigl\{\mid e^{2\omega -
\int A_e}\mid\bigr\} \sqrt{1+\xi\bar\xi}\vert\int^\xi_0{\rho^\prime(
\xi^\prime,\bar\xi)\over(1+\xi^\prime\bar\xi)^2}d\xi^\prime\vert=\cr
&=\const {\mid\xi\mid\over(1+\xi\bar\xi)\sqrt{1+\xi\bar\xi}},}\eqno(2.17)
$$

\noindent which has a zero in the north and the south poles of $\$ $. Hence
the anti-holomorphic spinor field $\lambda^0_R$ has a single zero, implying
that although $\dim\ker{\cal H}^-=2$ and $\$ $ is not future marginally
trapped, it is exceptional. \par
       Finally we note that Jeffryes's construction [18] can be repeated to
show that there might be 2-spheres on which the adjoint equation (2.8) has
at least one nontrivial solution; implying that for such surfaces $\dim\ker
{\cal H}^-\ge3$. For the existence of the nontrivial solution of (2.8) in
Jeffryes's construction, however, the vanishing both of $\sigma$ and
$\rho^\prime$ and some of their derivatives at least at two different
points is needed. Thus the strict positivity of $\rho^\prime$ on $\$ $
excludes all the three forms of exceptonal 2-spheres considered here. The
`strict convexity' of $\$ $ seems therefore to ensure the genericity of
$\$ $. This proposition is however not yet proved.\par
      If $\$ $ is generic then $\ker{\cal H}^\pm$ are therefore two
dimensional complex vector spaces and $\varepsilon^{\uA\uB}:=\varepsilon
^{RS}\lambda^{\uA}_R\lambda^{\uB}_S$ is a naturally defined constant
symplectic inner product. Then $\varepsilon^{\uA\uB}$ is invertible and
the space of the holomorphic/anti-holomorphic spinor fields is an $SL(2,
{\bf C})$-spinor space. In Minkowski spacetime the restriction to $\$ $
of the constant spinor fields (i.e. the spinor constituents of the
restriction to $\$ $ of the translation Killing vectors) can thus be
recovered as the solutions of ${\cal H}^\pm\lambda=0$. Substituting the
holomorphic or anti-holomorphic spinor fields into (1.6) we obtain four
real integrals, while both $A_{\$}$ and $I_{\$}$ are identically zero. \par
       The operators ${\cal C}^\pm$ are {\it not} elliptic, and hence
although $\$ $ is compact $\dim\ker{\cal C}^\pm$ are not necessarily
finite. In fact, if for example ${\cal G}=0$ and $\rho^\prime=0$ on $\$ $
then by (A.9) $\lambda^0$ must be zero and hence $\dim\ker{\cal C}^-=0$ if
$\supp\sigma^\prime=\$ $, while $\dim\ker{\cal C}^-=\infty$ if $\supp
\sigma^\prime\not=\$ $. On the other hand if $\sigma^\prime=0$ then by
(A.11) there are two independent solutions for $\lambda^0$ and hence
$\dim\ker{\cal C}^-=2$ if $\supp\rho^\prime=\$ $ (non-trapped round spheres,
for example), while $\dim\ker{\cal C}^-=\infty$ if $\supp\rho^\prime\not=
\$ $. Thus the operators ${\cal C}^\pm$ do not define acceptable spinor
propagation laws. The spinor fields belonging to $\ker{\cal C}^-$ are
precisely those satisfying $\Delta_a\lambda_R\pi^{+R}{}_K=0$. The
integrability conditon of this equation is

$$
\lambda^KF_{KRab}\pi^{+R}{}_S=-\Delta_e\lambda_K\pi^{-K}{}_L\Bigl(\delta^e
_aQ^L{}_{BB^\prime R}-\delta^e_bQ^L{}_{AA^\prime R}\Bigr)\pi^{+R}{}_S,
\eqno(2.18)$$

\noindent which for any pair $\lambda_R$, $\mu_R$ of spinor fields from $\ker
{\cal C}^-$ implies

$$
\Bigl(\lambda^R\mu^S\varepsilon_{RS}\Bigr)\pi^{-K}{}_CF^C{}_{Dab}\pi^{-D}
{}_L=\Delta_e\Bigl(\lambda^R\mu^S\varepsilon_{RS}\Bigr)\pi^{+K}{}_C\Bigl(
\delta^e_aQ^C{}_{BB^\prime D}-\delta^e_bQ^C{}_{AA^\prime D}\Bigr)\pi^{-D}
{}_L. \eqno(2.19)
$$

        The independent direct sums of $\Delta^\pm$ and ${\cal T}^\pm$
consisting of three terms are

$$
\eqalign{ {\cal J}^-:={\cal T}^-\oplus\Delta^+\oplus\Delta^-
 \hskip 10pt\approx & \hskip 10pt {\cal T}^-\oplus\Delta \hskip 10pt
 \approx \hskip 10pt {\cal H}^-\oplus\Delta^+\cr
{\cal J}^+:={\cal T}^+\oplus\Delta^+\oplus\Delta^-\hskip 10pt
 \approx & \hskip 10pt {\cal T}^+\oplus\Delta\hskip 10pt \approx
 \hskip 10pt {\cal H}^+\oplus\Delta^-\cr
{\cal K}^+:={\cal T}^+\oplus{\cal T}^-\oplus\Delta^+\hskip 10pt
 \approx &\hskip 10pt {\cal T}\oplus\Delta^+\hskip 10pt\approx\hskip 10pt
 {\cal T}^-\oplus {\cal H}^+\cr
{\cal K}^-:={\cal T}^+\oplus{\cal T}^-\oplus\Delta^-\hskip 10pt
 \approx &\hskip 10pt {\cal T}\oplus\Delta^-\hskip 10pt\approx\hskip 10pt
 {\cal T}^+\oplus{\cal H}^-.\cr}\eqno(2.20)
$$

\noindent Although they are not elliptic, they have finite dimensional
kernels. Since $\ker{\cal J}^\pm\subset\ker\Delta$ the operators ${\cal
J}^\pm$ cannot be used to define propagation laws. \par
      Since $\ker{\cal K}^\pm=\ker{\cal T}^\mp\cap\ker{\cal H}^\pm$ the
elements of $\ker{\cal K}^\pm$ are special
hol\-om\-orphic/anti-hol\-om\-orphic spinor fields, and hence for generic
2-spheres $\dim\ker{\cal K}^\pm\le 2$. Then if there were two
independent spinor fields in $\ker{\cal K}^-$, say $\lambda^R$ and
$\mu^R$, then $\lambda^R\mu^S\varepsilon_{RS}$ would be a nonzero constant
on $\$ $, and hence by (2.19) and (2.20) $\pi^{-K}{}_CF^C{}_{Dab}
\pi^{+D}{}_L$ would have to be zero. Thus for generic topological 2-spheres
$\dim\ker{\cal K}^\pm\leq 1$ and ${\cal K}^\pm$ do not yield the appropriate
number of quasi-local integrals.\par
       Finally the direct sum of all the irreducible chiral operators is

$$
{\cal C}:={\cal T}^+\oplus{\cal T}^-\oplus\Delta^+\oplus\Delta^-
\hskip 10pt \approx\hskip 10pt {\cal T}\oplus\Delta \hskip 10pt \approx
\hskip 10pt {\cal H}^+\oplus{\cal H}^-.\eqno(2.21)
$$

\noindent The spinor field $\lambda_R$ is holomorphic {\it and}
antiholomorphic iff it is $\Delta_a$--constant, which is equivalent to
$\Delta_{R^\prime (R}\lambda_{S)}=0$ by eqs.(6.9-12) and (6.1) of [1]. Thus
the elements of the kernel space of ${\cal C}$ are precisely the $\Delta_b
$-constant spinor fields on $\$ $. If $\lambda_R$ is a $\Delta_b$--constant
spinor field on $\$ $ then by (4.1) of [1] $\lambda^AF_{ABcd}=0$, which is
obviously equivalent to

$$
\lambda^AiF_{ABcd}t^ev^f\varepsilon_{ef}{}^{cd}=\lambda^A\Bigl(\psi
_{ABCD}\gamma^{CD}-\phi_{ABA^\prime B^\prime}\bar\gamma^{A^\prime
B^\prime}+2\Lambda\gamma_{AB}\Bigr)=0.\eqno(2.22)
$$

\noindent Thus $\$ $ admits a $\Delta_b$--constant spinor field $\lambda
_R$ only if the two principle spinors of $F_{ABcd}t^ev^f\varepsilon_{ef}
{}^{cd}$ coincide and are proportional to $\lambda_R$; i.e. in
algebraically general spacetimes $\dim\ker{\cal C}=0$. If $\lambda^0_R$ and
$\lambda^1_R$ are $\Delta_b$--constant spinor fields then $\lambda^0_R
\lambda^1_S\varepsilon^{RS}$ is constant on $\$ $; and hence either
$\lambda^1_R=c\lambda^0_R$ for some nonzero $c\in{\bf C}$ or $\lambda^0_R$
and $\lambda^1_R$ form a basis in the spinor spaces at each point of $\$ $.
If $\lambda^{\uA}_R$, $\uA=0,1$, are independent $\Delta_b$--constant
spinor fields then for any spinor field $\lambda_R$ there are functions
$\alpha$ and $\beta$ on $\$ $ such that $\lambda_R=\alpha\lambda^0_R+
\beta\lambda^1_R$, and if $\lambda_R$ is $\Delta_b$--constant then
$\alpha$ and $\beta$ are constant. Thus there are {\it at most} two
independent $\Delta_b$--constant spinor fields on $\$ $, when by (2.22)
$F_{ABcd}=0$ on $\$ $. In a {\it pp}-wave spacetime the constituent spinor
field $\lambda_A$ of the constant null vector can always be chosen to be
constant; i.e. to satisfy $\nabla_b\lambda_A=0$. Its restriction to $\$ $
is $\Delta_b$--constant and hence in a {\it pp}-wave spacetime
$\dim\ker{\cal C}\ge 1$. In Minkowski spacetime there are two linearly
independent constant spinor fields whose restriction to $\$ $ are the two
independent $\Delta_b$--constant spinor fields and $\dim\ker{\cal C}=2$.
In the next section we will show that the converse of these statements is
also true, namely assuming $\$ $ is a generic topological 2-sphere
bounding a spacelike hypersurface $\Sigma $ on which the dominant energy
condition holds and $\$ $ is `convex', the existence of one/two $\Delta
_b$--constant spinor field(s) on $\$ $ implies that the Cauchy development
$D(\Sigma)$ of $\Sigma$ is a {\it pp}-wave/flat spacetime geometry (and
hence in a nonflat {\it pp}-wave spacetime $\ker{\cal C}$ is {\it
precisely} 1 dimensional). \par
        To summarize our results on the kernel spaces of the first order
operators we have the following theorem:\par

\medskip \noindent
{\vbox {\bf Theorem 2.23:} The only first order linear differential
operators on $C^\infty(\$,{\bf S}_A)$ that are constructed only from the
chiral irreducible parts of the 2 dimensional Sen operator and have
generically 2 dimensional kernels are the holomorphy and anti-holomorphy
operators ${\cal H}^\pm$.} \par
\bigskip

\noindent The `natural' propagation laws for $\lambda_R$ are therefore
$\lambda\in\ker{\cal H}^\pm$. With this choice in the generic case we
have four real quasi-local Nester--Witten integrals and there is some hope
to obtain reasonable energy-momentum expressions. In fact, this is
precisely the Dougan--Mason energy-momentum [5]. In this framework it does
not seem to be possible to find quasi-local angular momentum expressions.
\par \vskip 1truecm

{\lbf 3. Quasi-local energy-momentum}\par \vskip 0.5truecm

\noindent  As we mentioned above the quasi-local energy-momentum suggested
by the covariant spinor formalism is precisely that proposed by Dougan and
Mason [5]. Their choice for the propagation laws was somewhat heuristic.
We can see, however, that among the first order propagation laws determined
by the Sen operator alone essentially this is the only possible choice. One
of the most important properties of the Dougan--Mason energy-momentum is
the mass-nonnegativity. This was proved in the GHP formalism, thus it might
be interesting first to see this proof in the covariant spinor formalism.
\par
       First consider spinor fields belonging only to $\ker\Delta^-$.
Suppose that $\$ $ is the boundary of a smooth spacelike 3 dimensional
submanifold $\Sigma$. Let $t^a$ be its future directed unit timelike
normal, $h_{ab}:=g_{ab}-t_at_b$ the induced metric, ${\cal D}_a:=h_a{}^b
\nabla_b$ the 3 dimensional Sen operator on $\Sigma$ [23] and
$\tilde\lambda_R$ be the solution of the Sen--Witten equation ${\cal
D}_{R^\prime}{}^R\tilde\lambda_R=0$ with an as yet unspecified boundary
condition on $\$ $. Then by the Reula--Tod form [14] of the 3 dimensional
Sen--Witten identity, which is just the pull back of the Sparling equation
(1.4) along the natural imbedding $i:\Sigma\rightarrow M$, we have

$$
H_{\$}[\tilde\lambda,{\overline{\tilde\lambda}}]={2\over\kappa}\int
_\Sigma\Bigl\{-h^{ab}t^{RR^\prime}\bigl({\cal D}_a\tilde\lambda_R
\bigr)\bigl({\cal D}_b{\overline{\tilde\lambda}}_{R^\prime}\bigr)-{1\over
2}\tilde\lambda^A{\overline{\tilde\lambda}}{}^{A^\prime}G_{ab}t^b
\Bigr\}d\Sigma,\eqno(3.1)
$$

\noindent where $d\Sigma$ is the induced volume element on $\Sigma$.
Thus if the dominant energy condition holds on $\Sigma $ then $H_{\$}[
\tilde\lambda,{\overline{\tilde\lambda}}]\ge 0$. On the other hand,
using $\Delta^-_{R^\prime}{}^R\lambda_R=0$ and equations (3.4), (4.4),
(4.6) and (5.3) of [1]

$$
\eqalign{H_{\$}[\lambda,{\overline\lambda}]-H_{\$}[\tilde\lambda,{\overline
{\tilde\lambda}}]={2\over\kappa}\oint_{\$}\Bigl\{&-Q^R{}_{RB^\prime B}
\bigl(\tilde\lambda^B-\lambda^B\bigr)\bigl({\overline{\tilde\lambda}}
{}^{B^\prime}-{\overline\lambda}{}^{B^\prime}\bigr)+\cr
&+\bigl({\overline{\tilde\lambda}}{}^{R^\prime}-{\overline\lambda}
{}^{R^\prime}\bigr)\bar\pi^{-K^\prime}{}_{R^\prime}\Delta_{K^\prime K}
\tilde\lambda^K+\bigl(\tilde\lambda^R-\lambda^R\bigr)\pi^{-K}{}_R\Delta_{
KK^\prime}{\overline{\tilde\lambda}}{}^{K^\prime}\Bigr\}d\$. \cr}
$$

\noindent The most natural boundary condition to ${\cal D}_{R^\prime}
{}^R\tilde\lambda_R=0$ would therefore be $\tilde\lambda^R\mid_{\$}=
\lambda^R$, which would ensure the non-negativity of $H_{\$}[\lambda,\bar
\lambda]$ too. The Sen--Witten equation, however, does not have in general
a solution on $\Sigma$ with this boundary condition. We should therefore
relax this boundary condition, and it seems natural next to choose $\pi
^{-S}{}_R(\tilde\lambda^R\mid_{\$}-\lambda^R )=0$. With this choice we have

$$
H_{\$}[\lambda,{\overline\lambda}]-H_{\$}[\tilde\lambda,{\overline{\tilde
\lambda}}]=-{2\over\kappa}\oint_{\$}Q^R{}_{RB^\prime B}\bar\pi^{+
B^\prime}{}_{A^\prime}\pi^{+B}{}_A\bigl({\overline{\tilde\lambda}}
{}^{A^\prime}-\bar\lambda^{A^\prime}\bigr)\bigl(\tilde\lambda^A -
\lambda^A\bigr)d\$.\eqno(3.2)
$$

\noindent If, following Dougan and Mason, we assume that the outgoing null
geodesics orthogonal to $\$ $ are not contracting on $\$ $, i.e.
$\rho^\prime\ge 0$, then, because of (4.9) of [1], this integral is
nonnegative. Furthermore $\rho^\prime\ge 0$ and the dominant energy
condition on $\Sigma$ ensure the existence of a solution $\tilde\lambda^R$
to the Sen--Witten equation with the boundary condition above [5]. Thus
$H_{\$}$ is a nonnegative hermitian scalar product on $\ker\Delta^-$, and
hence $H_{\$}$ satisfies the Cauchy--Schwartz inequality:

$$
H_{\$}[\lambda,\bar\lambda] H_{\$}[\mu,\bar\mu]\ge
H_{\$}[\lambda,\bar\mu] H_{\$}[\mu,\bar\lambda]\eqno(3.3)
$$

\noindent for any $\lambda_R$, $\mu_S\in\ker\Delta^-$. Similarly, the
dominant energy condition and $\rho\leq 0$ ensure the non-negativity of
$H_{\$}$ on $\ker\Delta^+$. Thus all the quasi-local energy-momentum
expressions in which $\lambda_R\in\ker\Delta^-$ or $\lambda_R\in\ker\Delta
^+$ is a part of the complete propagation law have this non-negativity
property. Such are for example the Ludvigsen--Vickers [3], the
Dougan--Mason [5] and the Bergqvist [6] propagation laws.\par
        The quasi-local energy-momentum is defined to be an element of the
dual space to the vector space of the `quasi-translations' of $\$ $.
Explicitly, if $\$ $ is generic and $\{\lambda_R^{\uA}\}$ is a basis in
$\ker{\cal H}^\mp$ then for any constant hermitian matrix $K_{\uA
\uB^\prime}$ the vector field $K_a:=K_{\uA \uA^\prime}\lambda^{\uA}_A
\bar\lambda^{\uA^\prime}_{A^\prime}$ can be interpreted as a
`quasi-translation' of $\$ $. These `quasi-translations' form a four real
dimensional subspace of $\ker{\cal H}^\mp\otimes{\overline {\ker{\cal
H}^\mp}}$ and span the four dimensional tangent spaces at the points of $\$ $.
Then the Dougan--Mason quasi-local energy-momentum is defined by $K_aP_{
\$}^a:=K_{\uA\uB^\prime}H_{\$}[\lambda^{\uA},\bar\lambda^{\uB^\prime}]$;
i.e. if $\{\lambda^{\uA}_R\}$ is a {\it normalized} spinor basis in $\ker
{\cal H}^\mp$ then the components of the quasi-local energy-momentum, the
quasi-local energy and mass are defined by

$$
\eqalignno{P_{\$}^{\uA\uB^\prime}&:=H_{\$}[\lambda^{\uA},\bar\lambda^{
\uB^\prime}],&(3.4)\cr
E_{\$}&:={1\over\sqrt2}\bigl(P_{\$}^{00^\prime}+
P_{\$}^{11^\prime}\bigr),&(3.5)\cr
m^2_{\$}&:=\varepsilon_{\uA\uB}\varepsilon_{\uA^\prime\uB^\prime}P_{\$}^{
\uA\uA^\prime}P_{\$}^{\uB\uB^\prime}=2\bigl(P_{\$}^{00^\prime}P_{\$}^{
11^\prime}-P_{\$}^{01^\prime}P_{\$}^{10^\prime}\bigr),&(3.6)\cr}
$$

\noindent respectively. Thus if the dominant energy condition holds and
in the anti-holomorphic case $\rho^\prime\ge 0$ on $\$ $ (and $\rho\leq
0$ on $\$ $ in the holomorphic case) then $E_{\$}\ge 0$ [5]
and by the Cauchy--Schwartz inequality $m^2_{\$}\ge 0$; i.e. $P_{\$}^a$
is a future directed nonspacelike vector [11]. This energy-momentum gives
the correct, expected value in the weak field approximation [5]. The
quasi-local energy-momentum is calculated for round spheres and small
and large spheres [17] and at the horizon of the Reissner--Nordstr\"om
and Kerr spacetimes [24] and compared with other definitions. At future
null infinity the quasi-local energy-momentum defined by the
anti-holomorphic spinor fields tends to the Bondi--Sachs four-momentum.
The expression based on the holomorphic spinor fields in general tends to
infinity. That yields the Bondi--Sachs four-momentum at {\it past} null
infinity [17]. At spacelike infinity both definitions give the ADM
energy-momentum.\par
     Recently it has been shown that for generic $\$ $ $P_{\$}^{\uA
\uB^\prime}=0$ iff the Cauchy development of $\Sigma$ is flat; but the
vanishing of the mass alone does not imply flatness. The zero-mass Cauchy
developments are precisely the {\it pp}-wave geometries with pure
radiation [11]. In the rest of this section only the anti-holomorphic
expression will be examined and we prove two theorems which give further
equivalent statements for the zero energy-momentum and zero mass spacetime
configurations. \par

\medskip\noindent
{\vbox {\bf Theorem 3.7:} Let $\$ $ be a generic 2-sphere for which
    $\rho^\prime\ge 0$, let $\Sigma$ be a spacelike hypersurface such that
    $\partial\Sigma=\$ $ and let the dominant energy condition hold on
    $\Sigma$. Then the following statements are equivalent:
\item{1.} $P^{\uA\uB^\prime}_{\$}=0$,
\item{2.} $E_{\$}=0$,
\item{3.} $D(\Sigma)$, the Cauchy development of $\Sigma$, is flat,
\item{4.} There exist two linearly independent $\Delta_e$--constant spinor
          fields on $\$ $.} \par

\medskip
\noindent {\it Proof}: 1. obviously implies 2. Since $H_{\$}$ is
non-negative, $E_{\$}=0$ implies both $H_{\$}[\lambda^0,\bar\lambda^{
0^\prime}]=0$ and $H_{\$}[\lambda^1,\bar\lambda^{1^\prime}]=0$.
They, as it was shown in [11], imply the flatness of $D(\Sigma)$. If
$D(\Sigma)$ is flat then, at least in an open neighbourhood of $\$ $ in
$D(\Sigma)$, there exist two linearly independent $\nabla_e$--constant
spinor fields. Their restriction to $\$ $ are independent $\Delta_e
$--constant spinor fields on $\$ $. If $\lambda^0_R$ and $\lambda^1_R$ are
$\Delta_e$--constant on $\$ $ then by

$$
\lambda^0_R,\lambda^1_R\in\ker\Bigl(\Delta^+\oplus\Delta^-\oplus
{\cal T}^+\oplus{\cal T}^-\Bigr)\subset\ker\Bigl(\Delta^+\oplus\Delta^-
\Bigr)=\ker H_{\$}
$$

\noindent $H_{\$}[\lambda^0,\bar\lambda^{0^\prime}]=H_{\$}[\lambda^1,
\bar\lambda^{1^\prime}]=0$, which by the Cauchy--Schwartz inequality imply
$P^{\uA\uB^\prime}_{\$}=0$.\par
\bigskip

     The equivalence of 1. and 3. has been discussed in [11], thus we
discuss only the equivalence of 2. and 3. and of 3. and 4. In the
(classical and quantum) theory of fields one can define the vacuum state
as the minimal energy state of the system and the ground state as in which
all the particle fields and field strength (of the gauge fields) are zero.
These states do not necessarily coincide even if the energy functional is
bounded below, as for example in the $\phi^4$--theory. The strict
positivity of the ADM and Bondi--Sach masses [14,15] implies that the
ground state (i.e. the flat spacetime) is the minimal energy state
among the states describing asymptotically flat spacetimes. This however
does not necessarily exclude the existence of non-asymptotically flat
spacetimes with negative quasi-local energy somewhere. Having accepted the
Dougan--Mason energy-momentum as `the' correct gravitational
energy-momentum, we can define quasi-locally the vacuum state of Einstein's
theory by $E_{\$}=0$. Then the equivalence of the statements 2. and 3.
thus means that the (quasi-locally defined) vacuum state is the uniquely
determined ground state, and hence no spontaneous symmetry breaking can
occur in Einstein's theory.\par
     The fact whether there exist two independent $\Delta_e$--constant
spinor fields on $\$ $ depends only on the 2 dimensional Sen--geometry of
$\$ $. On the other hand the equivalence of 3. and 4. of Theorem 3.7 means
that gravitation together with matter fields satisfying the dominant
energy condition is so `rigid' a system that the information that
$D(\Sigma)$ is flat is completely encoded into the Sen--geometry of $\$ $.
In other words flat Cauchy developments of a finite Cauchy surface
can be characterized not only by the usual Cauchy data on a {\it three
dimensional} $\Sigma$ but by the Sen--geometry of a spacelike {\it two
dimensional} sphere too.\par

\medskip\noindent
{\vbox {\bf Theorem 3.8:} Under the conditions of Theorem 3.7 the
   following statements are equivalent:
\item{1.} $m^2_{\$}=0$,
\item{2.} $D(\Sigma)$ is a {\it pp}-wave geometry with pure radiation; i.e.
      there exists a constant nonzero null vector field $L^a$ on $D(\Sigma)$
      such that $L^aT_{ab}=0$,
\item{3.} There exists a $\Delta_e$--constant spinor field on $\$ $.}\par

\medskip
\noindent {\it Proof}: The fact that 1. implies 2. was proved in [11]. If
there is a future directed constant nonzero null vector field $L^a$ on
$D(\Sigma)$ then there exists a $\nabla_e$--constant spinor field $\lambda
^A$ on $D(\Sigma)$ such that $L^a=\lambda^A\bar\lambda^{A^\prime}$. The
restriction of $\lambda^A$ to $\$ $ is a nonzero $\Delta_e$--constant
spinor field on $\$ $. If $\lambda^A$ is $\Delta_e$--constant on $\$ $ then
by

$$
\lambda_R\in\ker\Bigl(\Delta^+\oplus\Delta^-\oplus{\cal T}^+
\oplus{\cal T}^-\Bigr)\subset\ker H_{\$}
$$

\noindent $H_{\$}[\lambda,\bar\lambda]=0$. However $\lambda_R$ can be
chosen to be $\lambda^0_R$, one of the normalized basis spinors in
$\ker{\cal H}^-$. But then $P^{00^\prime}_{\$}=0$, which by the
Cauchy--Schwartz inequality implies $m^2_{\$}=0$.\par
\bigskip

      The equivalence of 1. and 2. was discussed in [11], thus we
consider only the equivalence of 2. and 3. Again, the existence of a
$\Delta_e$--constant spinor field depends only on the two dimensional
Sen--geometry of $\$ $. The equivalence of the statements 2. and 3. of
Theorem 3.8, on the other hand, means that the information that
$D(\Sigma)$ is a {\it pp}-wave geometry with pure radiation is completely
encoded into the Sen--geometry of $\$ $. There is, however, an essential
difference between the zero-energy and zero-mass cases. Namely while in the
zero-energy case we could determine the {\it metric} of $D(\Sigma)$, that
is flat, in the zero-mass case we can determine only the {\it class} of
the metric of $D(\Sigma)$: that is {\it pp}-wave plus pure radiation. Thus
naturally arises the question whether all the information on the metric of
$D(\Sigma)$ itself are encoded into the Sen--geometry of $\$ $. The answer
obviously depends on the detailes of the field equations for the matter
fields. For vacuum the answer is affirmative as there is a smooth function
$\Phi:\$\rightarrow{\bf C}$ whose second Sen--derivatives determine
completely the geometry of $D(\Sigma)$. $\Phi$ is constant iff
$D(\Sigma)$ is flat. The (vacuum) {\it pp}-wave Cauchy developments can
therefore be characterized not only by the usual Cauchy data on a {\it
three dimensional} hypersurface $\Sigma$ but by the {\it two dimensional}
Sen--geometry of $\$ $. The details of this analysis will be published in
a separate paper.\par
       If $\$ $ is exceptional and $\dim\ker{\cal H}^-=2$ (e.g. if $\$ $ is
future marginally trapped) then the quasi-translations are null and are all
proportional with each other, but the components $P_{\$}^{\uA\uB^\prime}$
with respect to a basis $\{\lambda^{\uA}_R\}$ of $\ker{\cal H}^-$ can
still be defined by (3.4). One can, however, see from (1.6) and Lemma
2.9.i that for the physically important special case of future
marginally trapped surfaces $P_{\$}^{\uA\uB^\prime}$ are zero. But in
this case $\varepsilon^{\uA\uB}:=\varepsilon^{RS}\lambda^{\uA}_R
\lambda^{\uB}_S$ is singular, thus one might conjecture that the
vanishing of $P_{\$}^{\uA\uB^\prime}$ does not mean the vanishing of the
quasi-local four-momentum, and the quasi-local mass of the marginally
trapped surfaces can be defined in a limiting procedure. In fact,
the quasi-local mass was calculated for round spheres and it was found that
the quasi-local mass has a well defined and nonzero limit even if the round
spheres tend to a marginally trapped surface [17]. Furthermore, the
thermodynamical analysis shows that a positive mass is associated
with the marginally trapped surfaces [25], and in general the irreducible
mass $({4\pi\over\kappa^2}\Area(\$))^{1\over2}$ is expected (see for example
[26]). Here we show that although for the family $\$(u)$ of 2-surfaces
considered in section 2 $m^2_{\$(u)}$ has a well defined positive limit,
the limiting value {\it does} depend on the family $\$(u)$.\par
       The solution of (2.11) with $\lambda^1_0=i({\nu\over\bar P})
^{1\over2}\bar\xi$ is

$$
\dot\lambda^1_1(\xi,\bar\xi)=-ie^{-{1\over2}\omega+{1\over2}\int A_e}
\sqrt{1+\xi\bar\xi}\Bigl\{c+\int_0^\xi {\dot\rho^\prime(\xi^\prime,\bar\xi)
\over(1+\xi^\prime\bar\xi)^2}\bar\xi e^{2\omega-\int A_e}d\xi^\prime
\Bigr\},\eqno(3.9)
$$

\noindent where the constant $c$ is the value of

$$
c(\xi,\bar\xi):=-\int^\xi_0{\dot\rho^\prime(\xi^\prime,\bar\xi)\over
(1+\xi^\prime\bar\xi)^2}\bar\xi e^{2\omega -\int A_e}d\xi^\prime
$$

\noindent in the north pole of $\$ $. Then by (2.14) and (3.9) the
derivative of $\Lambda(u):=\varepsilon^{RS}\lambda^0_R(u)\lambda^1_S(u)$
with respect to $u$ at $u=0$, given by (2.13), is just the constant $c$.
Let $\$(u)$ be generic for any nonzero $u$ (e.g. if $\dot\rho^\prime >0$
on $\$ $). Then $c\not=0$, $\varepsilon^{\uA \uB}(u)$ is invertable and
$m^2 _{\$(u)}$ can be defined (see (3.6)):

$$
m^2_{\$(u)}=2\Bigl( {P^{00^\prime}_{\$(u)}\over\Lambda(u)}{P^{11^\prime}
_{\$(u)}\over\bar\Lambda(u)}-{P^{01^\prime}_{\$(u)}\over\Lambda(u)}
{P^{10^\prime}_{\$(u)}\over\bar\Lambda(u)}\Bigr).\eqno(3.10)
$$

\noindent But by the L'Hospital rule and (2.11)

$$
\lim_{u\rightarrow 0}\Bigl({P^{{\uA}{\uB}^\prime}_{\$(u)}\over\Lambda(u)}
\Bigr)={2\over\kappa}{1\over c}\int_{\$}\dot\rho^\prime\lambda^{\uA}_0
\bar\lambda^{\uB^\prime}_{0^\prime}d\$,\eqno(3.11)
$$

\noindent and hence $\lim_{u\rightarrow 0}m^2_{\$(u)}$ does depend not
only on the geometry of $\$=\$(0)$ but on $\dot\rho^\prime$ too, i.e. on
the family of the 2-surfaces $\$(u)$ as well. In particular for the
2-sphere with $\rho^\prime=0$, $P={1\over\sqrt2 R}(1+\xi\bar\xi)$ and
$A_e=0$ the choice $\dot\rho^\prime=\const$ yields the expected value
$\lim_{u\rightarrow 0}m^2_{\$(u)}={4\pi\over\kappa^2}\Area(\$)$, while
the choice $\dot\rho^\prime={\const\over 1+\xi\bar\xi}$ yields ${32\pi
\over9\kappa^2}\Area(\$)$. \par

\vskip 1truecm

{\lbf Appendix: The $\edth $ operator}\par \vskip 0.5truecm

\noindent The aim of this appendix is to clarify the structure of the
kernel spaces of the edth operators if $\$ $ is homeomorphic to a 2-sphere.
First recall [7-10] that the spin connection on $(B,\$ ,{\bf C}^*)$
determines a connection on the associated vector bundle $E(p,q)$ of
scalars of weight $(p,q)$, $p-q\in {\bf Z}$. The corresponding covariant
directional derivations in the directions $m^a$ and $\bar m^a$ are the
usual edth operators $\edth_{(p,q)}$ and $\edth_{(p,q)}^\prime $ sending
smooth cross sections of $E(p,q)$ to smooth cross sections of $E(p+1,q-1)$
and $E(p-1,q+1)$, respectively. They are elliptic operators and their
index was calculated by Baston [10]:

$$
\eqalign{\index\edth_{(p,q)}&=\Bigl(1+p-q\Bigr)(1-{\cal G}),\cr
\index\edth_{(p,q)}^\prime&=\Bigl(1-p+q\Bigr)(1-{\cal G});\cr}\eqno(A.1)
$$

\noindent where ${\cal G}$ is the genus of the closed two-surface $\$ $,
and ${\cal G}=0$ for $\$ $ homeomorphic to a 2-sphere. The smooth section
$\phi\in E^\infty(p,q)$ is called anti-holomorphic with respect to
$\delta_e$ if $\edth_{(p,q)}\phi=0$. ($E^\infty(p,q)$ is the space of the
smooth cross sections of $E(p,q)$.) The multiplicity of the zeros of
anti-holomorphic sections can also be defined and the zeros are isolated.
Since for any $\phi\in E^\infty(p,q)$ and $\psi\in E^\infty(p^\prime,
q^\prime)$ $\phi\psi\in E^\infty(p+p^\prime,q+q^\prime)$ , the Leibnitz
rule $\edth_{(p+p^\prime,q+q^\prime)}(\phi\psi)=\psi\edth_{(p,q)}\phi+
\phi\edth_{(p^\prime,q^\prime)}\psi$ implies the inequality

$$
\dim\ker\edth_{(p+p^\prime,q+q^\prime)}\geq\max\Bigl\{\dim\ker\edth_{(p,q)},
\dim\ker\edth_{(p^\prime,q^\prime)}\Bigr\}\hskip 5pt
{\rm if}\hskip 5pt (\dim\ker\edth_{(p,q)})(\dim\ker\edth_{(p^\prime,
q^\prime)})\not= 0.\eqno(A.2)
$$

\noindent There is a similar inequality for the primed edth too. If $h$
is any fixed nowhere zero scalar of weight $(1,1)$ then for any $\phi,
\psi\in E^\infty(p,q)$

$$
\langle\phi,\psi\rangle_{(p,q)}:=\oint_{\$}\mid h\mid^{-(p+q)}\phi\bar\psi
d\$ \eqno(A.3)
$$

\noindent defines a hermitian inner product on $E^\infty(p,q)$. The adjoint
of $\edth_{(p,q)}$ and $\edth_{(p,q)}^\prime$ with respect to this inner
product is given by

$$
\eqalign{\Bigl(\edth_{(p,q)}\Bigr)^\dagger&=-\mid h\mid^{(p+q)}
\edth_{(-q+1,-p-1)}^\prime\mid h\mid^{-(p+q)},\cr
\Bigl(\edth_{(p,q)}^\prime\Bigr)^\dagger&=-\mid h\mid^{(p+q)}
\edth_{(-q-1,-p+1)}\mid h\mid^{-(p+q)};\cr}\eqno(A.4)
$$

\noindent respectively. Thus in general for $q\not=-p$ these adjoints,
and hence the kernel spaces $\ker(\edth_{(p,q)})^\dagger$ and $\ker(
\edth^\prime_{(p,q)})^\dagger$ depend on $h$. Since however (A.4) is a
similarity transformation between $(\edth_{(p,q)})^\dagger$ and $\edth
^\prime_{(-q+1,-p-1)}$ and $(\edth^\prime_{(p,q)})^\dagger$ and $\edth
_{(-q-1,-p+1)}$, respectively, we have

$$
\eqalign{\dim\ker\bigl(\edth_{(p,q)}\bigr)^\dagger&=\dim\ker\edth
^\prime_{(-q+1,-p-1)},\cr
\dim\ker\bigl(\edth^\prime_{(p,q)}\bigr)^\dagger&=\dim\ker\edth_{(-q-1,
-p+1)}.\cr}\eqno(A.5)
$$
\par
      In the rest of this appendix ${\cal G}=0$ will be assumed. Then
Liouville's theorem implies

$$
\dim\ker\edth_{(0,0)}=\dim\ker\edth_{(0,0)}^\prime=1.\eqno(A.6)
$$

\noindent By (A.1,5) this is equivalent to $\dim\ker\edth_{(-1,1)}=
\dim\ker\edth^\prime_{(1,-1)}=0$, which, by (A.1,2,5), implies
$\dim\ker\edth_{(1,-1)}=\dim\ker\edth_{(-1,1)}^\prime =3$. Using
(A.1,2,5,6) it is not difficult to show by induction that

$$
\eqalign{\dim\ker\edth_{(n,-n)}&=\dim\ker\edth_{(-n,n)}^\prime=2n+1\cr
\dim\ker\edth_{(-n,n)}&=\dim\ker\edth_{(n,-n)}^\prime=0\hskip 40pt
\forall n\in{\bf N}.\cr}\eqno(A.7)
$$

\noindent Since by (A.1) $\dim\ker\edth_{(p,p)}\ge1$ and $\dim\ker\edth
_{(p,p)}^\prime\ge 1$ $\forall p\in{\bf R}$, (A.2) and (A.6) imply

$$
\dim\ker\edth_{(p,p)}=\dim\ker\edth_{(p,p)}^\prime=1\hskip 10pt \forall
p\in{\bf R}.\eqno(A.8)
$$

\noindent Thus there is precisely one scalar $h_0\in E^\infty(1,1)$
satisfying $\edth h_0=0$ and precisely one scalar $h^\prime_0\in E^\infty
(-1,-1)$ satisfying $\edth^\prime h^\prime_0=0$ (and hence $h^\prime_0=
\bar h_0$). \par
       For any $p\in{\bf R}$ and $n\in{\bf N}$ (A.1) implies $\dim\ker
\edth_{(-p+n,-p)}\ge1+n$ and $\dim\ker\edth^\prime_{(-p,-p+n)}\ge1+n$.
Thus by (A.2) and (A.6) we have

$$
\dim\ker\edth_{(p-n,p)}=\dim\ker\edth^\prime_{(p,p-n)}=0 \hskip 10pt
\forall p\in{\bf R},\hskip 5pt n\in{\bf N}.\eqno(A.9)
$$

\noindent Similarly, (A.1) implies $\dim\ker\edth_{(p+2n,p)}\ge1+2n$ and
$\dim\ker\edth^\prime_{(p,p+2n)}\ge1+2n$ and then by (A.2), (A.7) and
(A.8)

$$
\dim\ker\edth_{(p+2n,p)}=\dim\ker\edth^\prime_{(p,p+2n)}=1+2n \hskip
10pt \forall p\in{\bf R},\hskip 5pt n\in{\bf N}.\eqno(A.10)
$$

\noindent Finally by (A.1),(A.5) and (A.9)

$$
\dim\ker\edth_{(p+2n-1,p)}=\dim\ker\edth^\prime_{(p,p+2n-1)}=2n\hskip
10pt \forall p\in{\bf R},\hskip 5pt n\in{\bf N}.\eqno(A.11)
$$

\noindent (A.8)-(A.11) is the complete list of the dimension of the kernel
spaces of the edth operators.\par
          For certain combinations of $p$ and $q$ the elements of these
kernel spaces can be realized by special vector and spinor fields on $\$ $.
First consider {\it surface} vector fields $K^a$ satisfying

$$
m^a\delta_aK^b=0\hskip 20pt\Bigl(\Pi^b_eK^e=K^b\Bigr);\eqno(A.12)
$$

\noindent i.e. that are anti-holomorphic with respect to the induced
Levi--Civit\`a connection, or, in other words, with respect to the
{\it intrinsic} complex structure of $\$ $. The components of $K^b$,
$\lambda:=m_bK^b$ and $\tilde\lambda:=\bar m_bK^b$, are scalars of weight
(1,-1) and (-1,1), respectively; and (A.12) is equivalent to $\edth_{(1,
-1)}\lambda=0$ and $\edth_{(-1,1)}\tilde\lambda=0$. Since $\dim\ker\edth_{
(-1,1)}=0$, $K^b$ satisfying (A.12) must be proportional to $\bar m^b$.
The explicit solutions of (A.12) can be given in a coordinate system. Let
$\{(U,\xi),(V,\eta)\}$ be the usual complex analitic atlas for $\$ $ (i.e.
$n\in\$ $ (`north pole'), $U:=\$-\{n\}$, $\xi:U\rightarrow{\bf C}$ a
homeomorphism, $s\in\$ $ such that $\xi(s)=0$ (`south pole'), $V:=\$-\{s\}$
and $\eta:V\rightarrow {\bf C}$ such that $\eta={1\over\xi}$ on $U\cap V$
and $\eta(n)=0$); and let the vector field $m^a$ be chosen to be
proportional to ${\partial\over\partial\xi}$: $m^a=P(\xi,\bar\xi)({\partial
\over\partial\xi})^a$ (see e.g. [8]). Then in $(U,\xi)$ the general
solution of (A.12) is a countable linear combination of the vector fields

$$
K^a_m:={\bar\xi}^m\bigl({\partial\over\partial\bar\xi}\bigr)^a,\hskip
20pt m=0,1,2,... \eqno(A.13)
$$

\noindent These are the independent conformal Killing vectors of $(U,q
_{ab})$, their Lie algebra is the Virasoro algebra (without the central
extension): $[K_m,K_n]^a=(n-m)K^a_{m+n-1}$, and $k^a_z:=iK^a_1$, $k^a_y:=
-{1\over\sqrt2}(K^a_0+{1\over2}K^a_2)$, $k^a_x:={i\over\sqrt2}(K^a_0-
{1\over2}K^a_2)$ are the usual generators of its $so(3,{\bf C})$
subalgebra. The behaviour of the vector fields $K^a_m$ in the `north pole'
$n$ can be clarified in $(V,\eta)$: Only $K^a_0$, $K^a_1$ and $K^a_2$ are
well defined on the whole $\$ $ and $K^a_0$ has a zero in $n$ (of
multiplicity 2), $K^a_1$ has zeros in $n$ and $s$ (both of multiplicity 1)
and $K^a_2$ has a zero in $s$ (of multiplicity 2). The corresponding (1,-1)
weight scalars (in the $(U,\xi)$ coordinate system): $\lambda_0=-\bar
P^{-1}$, $\lambda_1=-\bar P^{-1}\bar\xi$ and $\lambda_2=-\bar P^{-1}
\bar\xi^2$, respectively. They are the independent elements of $\ker\edth_{
(1,-1)}$ and obviously $\lambda_0\lambda_2=(\lambda_1)^2$. \par
        Next consider {\it normal} vector fields $N^a$ satisfying

$$
m^a \delta _a N^b=0 \hskip 20pt\Bigl(\Pi ^b_eN^e=0\Bigr);\eqno(A.14)
$$

\noindent i.e. which are anti-holomorphic with respect to the induced
Levi--Civit\`a connection. Then $\nu:=l_aN^a$ and $\tilde \nu:=n_aN^a$ are
scalars of weight (1,1) and (-1,-1), respectively; and (A.14) is equivalent
to $\edth_{(1,1)}\nu=0$ and $\edth_{(-1,-1)}\tilde\nu=0$. Since $\dim\ker
\edth_{(1,1)}=\dim\ker\edth_{(-1,-1)}=1$, (A.14) has precisely two
independent solutions: $\nu n^b$ and $\tilde\nu l^b$; furthermore if
$\edth_{(1,1)}\nu=0$ then, provided $\nu$ is nowhere zero, $\tilde\nu=
\nu^{-1}$. In the coordinate system $(U,\xi)$ above (A.14) takes the form

$$
P{\partial \nu\over\partial\xi}+\nu A_em^e=0,\eqno(A.15)
$$

\noindent where $A_e$ is the `boost gauge potential'. The general
solution of (A.15) on $U$ is the countable linear combination of the scalars

$$
\nu_n:=\bar\xi^n\exp\Biggl(-\int^\xi_{\xi_0}A_e\Biggr),\hskip
20pt n=0,1,2,... \eqno(A.16)
$$

\noindent Here the integration is taken along the complex path whose
tangent is $m^a$. However $\nu_n n^b$ is well defined on the whole $\$ $
only for $n=0$. This serves the only independent element of $\ker\edth_{
(1,1)}$, and since this is nowhere zero, the only independent element of
$\ker\edth_{(-1,-1)}$ too. \par
        Finally consider spinor fields satisfying

$$
m^a\delta _a \omega^R=0; \eqno(A.17)
$$

\noindent i.e. that are anti-holomorphic with respect to the induced
Levi--Civit\`a connection. $\omega:=\omega^Ro_R$ and $\tilde\omega:=
\omega^R\iota_R$ are scalars of weight (1,0) and (-1,0), respectively, and
(A.17) is equivalent to $\edth_{(1,0)}\omega=0$ and $\edth_{(-1,0)}\tilde
\omega=0$. Since however $\dim\ker\edth_{(-1,0)}=0$ and $\dim\ker\edth_{
(1,0)}=2$, there are two independent solutions of (A.17), and both must
have the form $\omega^R=\omega\iota^R$. Let $(o^A,\iota^A)$ be a local
cross section of the spin frame bundle $(B,\$,{\bf C}^*)$ on $U$ so that
$o^A\bar\iota^{A^\prime}=m^a=P(\xi,\bar\xi)({\partial\over\partial\xi})^a$,
and let $\omega_0$ and $\omega_1$ be the two independent solutions.
If $\nu$ is the solution of (A.15) then $\omega_0 ^2\nu^{-1}$,
$\omega_0\omega_1\nu^{-1}$ and $\omega_1^2\nu^{-1}$ are independent
elements of $\ker\edth_{(1,-1)}$, and hence they may be chosen to be
$\lambda_0$, $\lambda_1$ and $\lambda_2$ above, respectively. Therefore
on $U$

$$
\omega_0=i\bigl({\nu\over\bar P}\bigr)^{1\over2},\hskip 20pt
\omega_1=i\bigl({\nu\over\bar P}\bigr)^{1\over2}\bar\xi. \eqno(A.18)
$$

\noindent The corresponding spinor fields have single zeros of multiplicity
1 in the `north pole' $n$ and in the `south pole' $s$, respectively. These
serve the independent elements of $\ker\edth_{(1,0)}$, and $\{\omega^n_0,
\omega^{n-1}_0\omega_1,$ $...,$ $\omega^n_1\}$ is a basis in $\ker\edth_{
(n,0)}$. The sum of the multiplicity of the zeros of the elements of $\ker
\edth_{(n+p,p)}$ is precisely $n$ for any $n\in{\bf N}$ and $p\in{\bf R}$.
\par \vskip 1truecm

{\lbf Acknowledgement}\par
\vskip 0.5truecm

\noindent I am greatful to Dr. \'A. Sebesty\'en and Dr. K. Szlach\'anyi
for their help in clarifying the structure of the kernel spaces, and to
one of the referees of [11] for his illuminating remarks on marginally
trapped surfaces. Thanks are due to the Soros Foundation and the North
Carolina State University for the financial support during the L\'anczos
Conference, where a part of this paper could be presented. This work was
partially supported by the Hungarian Scientific Research Fund grant OTKA
1815.\par \vskip 1truecm

{\lbf References}\par
\vskip 0.5truecm

\item{[1]}  L.B. Szabados, Two dimensional Sen connections in general
            relativity, preprint, 1994
\item{[2]}  R. Penrose, Proc.Roy.Soc.Lond. {\bf A 381} 53 (1982)
\item{[3]}  M. Ludvigsen, J.A.G. Vickers, J.Phys.A: Math.Gen. {\bf 16} 1155
            (1983)
\item{[4]}  G. Bergqvist, M. Ludvigsen, Class.Quantum Grav. {\bf 6} L133
            (1988); G. Bergqvist, M. Ludvigsen, Class.Quantum Grav. {\bf 8}
	     L29 (1991)
\item{[5]}  A.J. Dougan, L.J. Mason, Phys.Rev.Lett. {\bf 67} 2119 (1991)
\item{[6]}  G. Bergqvist, Class.Quantum Grav. {\bf 9} 1917 (1992)
\item{[7]}  R. Geroch, A. Held, R. Penrose, J.Math.Phys. {\bf 14} 874 (1973)
\item{[8]}  R. Penrose, W. Rindler, Spinors and Spacetime, Vol.1, Cambridge
            Univ. Press, 1984
\item{[9]}  S.A. Hugget, K.P. Tod, An Introduction to Twistor Theory,
            (London Mathematical Society Texts 4) Cambridge University
	     Press, Cambridge 1985
\item{[10]} R.J. Baston, Twistor Newsletter, {\bf 17} 31 (1984)
\item{[11]} L.B. Szabados, Class.Quantum Grav. {\bf 10} 1899 (1993)
\item{[12]} R. Penrose, W. Rindler, Spinors and Spacetime, Vol.2, Cambridge
            Univ. Press, 1986
\item{[13]} L.J. Mason, J. Frauendiener, The Sparling 3-form, Ashtekar
            Variables and Quasi-local Mass, in {\it Twistors in Mathematics
            and Physics} (London Math. Soc. Lecture Notes No. 156) Ed.:
	     T.N. Bailey and R.J. Baston, Cambridge Univ. Press, Cambridge
	     1990
\item{[14]} O. Reula, K.P. Tod, J.Math.Phys. {\bf 25} 1004 (1984)
\item{[15]} M. Ludvigsen, J.A.G. Vickers, J.Phys.A: Math.Gen. {\bf 15} L67
            (1982)
\item{[16]} K.P. Tod, Penrose's Quasi-local Mass, in {\it Twistors in
            Mathematics and Physics} (London Math. Soc. Lecture Notes No.
	     156) Ed.: T.N. Bailey and R.J. Baston, Cambridge Univ. Press,
	     Cambridge 1990
\item{[17]} A.J. Dougan, Class.Quantum Grav. {\bf 9} 2461 (1992)
\item{[18]} B.P. Jeffryes, Class.Quantum Grav. {\bf 3} L9 (1986)
\item{[19]} A.D. Helfer, Class.Quantum Grav. {\bf 9} 1001 (1992)
\item{[20]} S. Kobayashi, K. Nomizu, Foundations of Differential Geometry,
            Vol 2, Interscience, 1968
\item{[21]} R.O. Wells, Differential Analysis on Complex Manifolds,
            Prentice--Hall, INC., Englewood Cliffs, New Jersey 1973
\item{[22]} R.C. Gunning, Lectures on Riemann Surfaces, (Princeton
             Mathematical Notes) Princeton Univ. Press, Princeton, New
	      Jersy 1966
\item{[23]} A. Sen, J.Math.Phys. {\bf 22} 1781 (1981)
\item{[24]} G. Bergqvist, Class.Quantum Grav. {\bf 9} 1753 (1992)
\item{[25]} R.M. Wald, The first law of black hole mechanics, University
            of Chicago preprint, 1993
\item{[26]} D. Christodoulou, S.-T. Yau, Some remarks on the quasi-local
            mass, in {\it Mathematics and General Relativity}, (Contemporary
	     Mathematics No 71) Ed.: J.A. Isenberg, AMS, New York 1988
\end